\magnification=\magstep1
\hfuzz= 6pt
\baselineskip=15pt
\hsize=6.0 truein

$ $
\vskip 1in
\centerline{{\bf A quantum-mechanical Maxwell's
demon}\footnote{$^\dagger$}{This work supported
in part by
grant \# N00014-95-1-0975 from the Office of Naval
Research, and by ARO and DARPA under grant \#
DAAH04-96-1-0386 to
QUIC, the Quantum Information and Computation
initiative.}}

\bigskip
\centerline{Seth Lloyd}

\centerline{d'Arbeloff Laboratory for Information
Systems and Technology}

\centerline{Department of Mechanical Engineering}

\centerline{Massachusetts Institute of Technology}

\centerline{MIT 3-160, Cambridge, Mass. 02139}

\centerline{slloyd@mit.edu}

\bigskip
\noindent {\it Abstract:} A Maxwell's demon is
a device that gets information
and trades it in for thermodynamic advantage, in apparent
(but not actual) contradiction to the second law of
thermodynamics. Quantum-mechanical versions of Maxwell's demon
exhibit features that classical versions do not: in particular,
a device that gets information about a quantum system disturbs
it in the process.  In addition, the information produced
by quantum measurement acts as an additional source of thermodynamic
inefficiency.  This paper investigates the properties of
quantum-mechanical Maxwell's demons, and proposes
experimentally realizable models of such devices.
\vfill\eject

\noindent{\bf Introduction}

In 1871, Maxwell noted that a being that could
measure the velocities of individual molecules in a gas
could shunt fast molecules into one container and slow
molecules into another, thereby creating a difference
in temperature between the two containers, in apparent
violation of the second law of thermodynamics.$^1$
Kelvin called this being a `demon': by getting
information and being clever how it uses it, such a
demon can in principle perform useful work.
Maxwell's demon has been the subject of considerable
discussion over the last century.$^2$
The contemporary view of the demon, spelled out in
the last decade,$^3$ is that a demon could indeed
perform useful work $k_B{\rm ln 2} T$
for each bit obtained, but must increase entropy by
at least $k_B {\rm ln} 2$ for each bit erased (a result known
as `Landauer's principle'$^4$).  As a result,
a demon that operates in cyclic fashion, erasing
bits after it exploits them, cannot violate the
second law of thermodynamics.

Up to now, Maxwell's demon has functioned primarily
as a {\it gedankenexperiment}, a thought experiment
that allows the exploration of theoretical issues.
This paper, in contrast, proposes a model of a
Maxwell's demon that could be realized experimentally using
magnetic or optical resonance techniques---all that
is required of a device to function as a quantum `demon' is
the ability to perform simple quantum logic operations.

Any experimentally realizable model of a `demon,'
like the molecules of Maxwell's original example,
must be intrinsically quantum-mechanical.
The classic reference on quantum
demons is Zurek's treatment of the quantum Szilard engine$^5$
(see also references 6-7).
There are compelling reasons to investigate quantum-mechanical
models of Maxwell's demon.  First,
as Zurek emphasized, a quantum-mechanical treatment
of the demon is necessary for making sense of the underlying
thermodynamics.  Second, the demon
operates by obtaining information, i.e., by making
measurements, and measurements on quantum systems
tend to disturb the system measured.  Third,
measurement of quantum systems
introduces `new' bits of information into the world.  As
noted by Haus$^8$, Landauer's principle implies that
these new bits come with a thermodynamic
cost: quantum demons suffer from peculiarly quantum sources
of inefficiency.

This paper presents a simple quantum-mechanical model
of a Maxwell's demon that allows these three issues
to be addressed.  The results are in accordance with
Zurek's treatment of the quantum Szilard engine:
quantum mechanics allows the consistent treatment
of thermodynamics, and introduces additional complications
because of the nature of quantum measurement.
The mechanisms by which information is
obtained, exploited, and erased are explored in detail.  Naturally,
such a device cannot violate the second law of thermodynamics,
but if supplied with heat reservoirs at different temperatures
it can undergo a Carnot cycle and function as a heat engine.
A thorough investigation of the thermodynamics of the
engine shows that Landauer's
principle is validated, and that each bit of information
introduced by quantum measurement and decoherence
functions as an extra bit of entropy, decreasing the
engine's efficiency.  A quantum-mechanical engine that
processes information is limited not by the Carnot
efficiency, but by the potentially lower quantum efficiency
$\varepsilon_Q$ defined below.

\bigskip\noindent{\bf 1. A magnetic resonance model
of a quantum `demon.'}

To see how a quantum system can function as a heat engine,
consider a spin in a magnetic field.  If the spin points
in the same direction as the field it has energy $-\mu B$,
where $\mu$ is the spin's magnetic dipole moment, and $B$ is
the field strength.  If it points in the opposite direction
it has energy $+\mu B$.  The state of the spin can be
controlled by conventional magnetic resonance techniques:
for example, the spin can be flipped from one energy
state to the other by applying a $\pi$ pulse
at the spin's Larmor precession
frequency $\omega = 2\mu B/ \hbar$.$^{9-10}$

When the spin flips, it exchanges energy with the oscillatory
field.  If the spin flips from the lower energy state to the
higher energy state it coherently absorbs one
photon with energy $\hbar\omega$ from the field; if
it flips from the higher energy state to the lower, it coherently
emits one photon with energy $\hbar\omega$ to the field.
The energy exchange involves no
entropy increase or loss of quantum coherence:$^{11}$ either the
field does work on the spin, or the spin does work on
the field.

It is clear how a device that acquires information
about such a spin could use the information to
make the spin do work.  Suppose that some
device can measure whether the spin is in the
low-energy quantum state $|\downarrow\rangle$ that points in
the same direction as the field, or in the high-energy
quantum state $|\uparrow\rangle$ that points in the opposite
direction to the field, and
if the spin is in the high-energy state,
send in a $\pi$-pulse to extract its energy.  The device
can then wait for the spin to come to equilibrium
at temperature $T_1>>2\mu B / k_B$ and repeat the
operation.  Each time it does so, it converts an average of
$\mu B$ of heat into work.  The device
gets information and uses that
information to convert heat into work.
The amount of work done by such a device operating
on a single spin is negligible;
but many such devices operating in parallel could
function as a `demon' maser, coherently amplifying
the pulse that flips the spins.

Landauer's principle prevents such a device from
violating the second law of thermodynamics.
To operate in a cyclic fashion, the device must
erase the information that it has gained about the
state of the spin.  When this information
is erased, entropy $S_{\rm out} \geq
k_B {\rm ln}2$ is
pumped into the device's environment, compensating
for the entropy $S_{\rm in} \approx k_B {\rm ln}2$ of
entropy in the spin
originally.  If the environment is a heat bath
at temperature $T_2$, heat $k_BT_2{\rm ln} 2$ flows
to the bath along with the entropy, decreasing
the energy available to convert into work.

The overall accounting of energy and entropy in the
course of the cycle is as follows:
heat in $Q_{\rm in}=T_1 S_{\rm in}$,
heat out $Q_{\rm out}=T_2 S_{\rm out}$,
work out $W_{\rm out}= Q_{\rm in} - Q_{\rm out}$,
efficiency
$$\varepsilon= W_{\rm out}/Q_{\rm in}
= 1-T_2 S_{\rm out}/T_1 S_{\rm in} \leq 1-T_2/T_1 \equiv
{\varepsilon}_C,\eqno(1)$$
\noindent where $\varepsilon_C$ is the Carnot efficiency.
Since $S_{\rm out}
\geq S_{\rm in}$, $W_{\rm out}$ can be greater than
zero only if $T_1>T_2$.  Landauer's principle implies
that instead of violating the second law of
thermodynamics, the device operates as a heat engine,
pumping heat from a high-temperature reservoir to a
low-temperature reservoir and doing work in the process.

Why quantum measurement introduces added
inefficiency into the operation of such a
device can be readily understood.
Suppose that the spin is originally in the state
$|\rightarrow\rangle=\allowbreak 1/{\sqrt 2}(|\uparrow\rangle +
|\downarrow\rangle)$.  One way to extract energy
from such a spin is to apply a $\pi/2$ pulse to
rotate the spin to the state $|\downarrow\rangle$,
extracting work $\mu B$ in the process.  A second
way to extract energy is to repeat the process
described above: measure the spin to see if it
is in the state $|\uparrow\rangle$, and if it
is, apply a $\pi$ pulse to extract work $2\mu B$.
This process also generates work $\mu B$ on
average, but in addition generates a `waste' bit of
information that costs energy $k_B T_2{\rm ln} 2$
to erase.  Quantum measurement introduces
added inefficiency to the process of getting
information about a quantum system
and exploiting that information to perform work.

More generally, suppose the spin is initially described
by a density matrix $\rho$.  The device makes a
measurement on the spin that takes
$$\rho\rightarrow \rho' = \sum_i P_i \rho P_i,\eqno(2)$$
\noindent where $P_i$ are
projection operators onto the eigenspaces of
the operator corresponding to the measurement.$^{12}$
The extra information generated by quantum
measurement is
$$\Delta S_Q/k_B {\rm ln}2 =
-{\rm tr} \rho'{\rm log}_2 \rho' -
(- {\rm tr} \rho{\rm log}_2 \rho)\geq 0\eqno(3)$$
\noindent The efficiency of the device
in converting heat to work is limited not by the
Carnot efficiency $\varepsilon_C$ but by the
quantum efficiency
$$\varepsilon_Q = 1-T_2(S_{\rm in}+
\Delta S_Q)/T_1S_{\rm in} =
\varepsilon_C -
T_2\Delta S_Q/T_1S_{\rm in}
.\eqno(4)$$
\noindent Equation (4) quantifies the inefficiency
due to any process, such as decoherence, that
destroys off-diagonal terms of the density matrix
$\rho$.$^{12-14}$

To investigate more thoroughly how
quantum measurement and decoherence introduce
thermodynamic inefficiency in a quantum
information-processing `demon' requires
a more detailed model of how such a device gets and
gets rid of information.
One of the simplest quantum systems that can function as a
measuring device is another spin.  Magnetic resonance affords
a variety of techniques, called spin-coherence double
resonance, whereby one spin can coherently
acquire information about another spin with which
it interacts.$^{9-10}$  The basic idea is
to apply a sequence of pulses that makes spin 2
flip if and only if spin 1 is in the excited state
$|\uparrow\rangle_1$, while leaving spin 1 unchanged.  If spin 2 is
originally in the ground state $|\downarrow\rangle_2$,
then after the conditional spin-flipping operation, the
two spins will either be in the state $|\uparrow\rangle_1
|\uparrow\rangle_2$ if spin 1 was originally in the state
$|\uparrow\rangle_1$, or in the state $|\downarrow\rangle_1
|\downarrow\rangle_2$ if spin 1 was originally in the state
$|\downarrow\rangle_1$.  Spin 2 has acquired information about
spin 1.  A variety of spin coherence double
resonance techniques (going under
acronyms such as {\it INEPT} and {\it INADEQUATE}) can be used
to perform this conditional flipping operation,$^{9-10}$
which can be thought of as an experimentally realizable
version of Zurek's treatment of the measurement process
in reference (5).  Readers familiar with quantum
quantum computation will recognize the conditional spin flip as
the quantum logic operation `controlled--$NOT$.'$^{15-17}$

How can this information be used to extract energy from spin
1?  Simply apply a second pulse sequence to
flip spin 1 if and only if
spin 2 is in the state $|\uparrow\rangle_2$, while leaving
spin 2 unchanged.  The energy transfer from spins to
field is as follows.  If spin 1 was originally in the state
$|\downarrow\rangle_1$, then spin 1 and spin 2 remain in the
state $|\downarrow\rangle$ through both pulses and no energy
is transferred to the field.  If spin 1 was originally in
the state $|\uparrow\rangle$, first spin 2 flips, then
spin 1, yielding a transfer of energy from spins to field of
$\hbar(\omega_1-\omega_2)=2(\mu_1-\mu_2)B$, which is $>0$ as
long as $\mu_1>\mu_2$.
The average energy extracted is half this
value.  As long as the conditional spin flips are performed
coherently, the amount of energy extracted depends only
on overall conservation of energy, and is independent of the
particular double resonance technique used.  Note that
the entire process maintains quantum
coherence and can be reversed simply by repeating the
conditional spin flips in reverse order.

\bigskip\noindent{\bf 2. A quantum heat engine}

To complete the treatment of the thermodynamics of this
device and to understand the role of decoherence and
quantum measurement in
its functioning, we must investigate how the `demon' interacts
with its thermal environment to take in heat and erase
information.  This section will show that a quantum device
that interacts with a thermal environment can indeed
get information and `cash it in' to do useful work, but
not by violating the second law of thermodynamics:
a detailed model of the erasure process confirms
Landauer's principle (one bit of information `costs'
entropy $k_B{\rm ln2}$).  As a result, instead of functioning as
a perpetual motion machine, the device operates as
a heat engine that undergoes a Carnot cycle.

The environment for our spins
will be taken to consist of two sets of modes of the
electromagnetic field, the first a set of modes at temperature $T_1$
with average frequency $\omega_1$ and with frequency spread
greater than $|\gamma|$ but less than $\omega_1-\omega_2$,
and the second a set of modes at temperature $T_2$
with average frequency $\omega_1$ and the same frequency spread.
Such an environment can be obtained, for example, by bathing
the spins in incoherent radiation with the given
frequencies and temperatures.  The purpose of such an
environment is to provide effectively separate heat
reservoirs for spin 1 and spin 2: spin 1 interacts strongly
with the on-resonance radiation at frequency $\omega_1$, and
weakly with the off-resonance radiation at
frequency $\omega_2$ --- {\it vice versa} for spin 2.
Over short times, to a good approximation
spin 1 can be regarded as interacting
only with mode 1, and spin 2 as interacting only with mode 2.
A spin can be put in and out of `contact' with its
reservoir by isentropically altering the frequency
of the reservoir mode to put the spin in and out of
resonance.

With this approximation, the initial probabilities
for the state of the $j$-th spin are (ignoring for the
moment the coupling between the spins)
$$p_j(\uparrow) = e^{-\mu_jB/k_BT_j}/Z_j,\quad
p_j(\downarrow) = e^{\mu_jB/k_BT_j}/Z_j, \eqno(5a)$$
\noindent yielding energy
$$E_j= -\mu_iB {\rm tanh}(\mu_j B/ k_B T_j), \eqno(5b)$$
\noindent and entropy
$$S_j= -k_B\sum_{i=\uparrow,\downarrow} p_j(i) {\rm ln} p_j(i)
= E_j/T_j + k_B{\rm ln} Z_j,\eqno(5c)$$
\noindent where $Z_j =
e^{-\mu_jB/k_BT_j} + e^{\mu_jB/k_BT_j}
= 2{\rm cosh} (\mu_jB/k_BT_j)$.
Even though it does not start out in a
definite state, spin 2 can still acquire information
about spin 1, and this information can be exploited to do
electromagnetic work.  The spins can function as a
heat engine by going through the following cycle:

\item{(1)} Using spin coherence double resonance, flip spin
2 iff spin 1 is in the state $|\uparrow\rangle_1$.  This
causes spin 2 to gain information
$(\tilde S_2-S_2)/k_B{\rm ln}2$
about spin 1 at the expense
of work $W_1=p_1(\uparrow) 2\mu_2B {\rm tanh}(\mu_2B/k_BT_2)$
supplied by the oscillating field.  Here $\tilde S_2=
-k_B\sum_{i=\uparrow, \downarrow} \tilde p_2(i)
{\rm ln} \tilde p_2(i)$, where
$\tilde p_2(\uparrow) = p_1(\uparrow) p_2(\downarrow)
+ p_1(\downarrow) p_2(\uparrow)$ and
$\tilde p_2(\downarrow) = p_1(\downarrow) p_2(\downarrow)
+ p_1(\uparrow) p_2(\uparrow)$ are the probabilities
for the states of spin 2 after the conditional spin flip.

\item{(2)} Flip spin
1 iff spin 2 is in the state $|\uparrow\rangle_1$.  This
step allows spin 2 to `cash in' $(S_2-S_1)/k_B{\rm ln}2$ of
the information it has acquired, thereby
performing work $-\mu_1B\big({\rm tanh}(\mu_1B/k_BT_1)
-{\rm tanh}(\mu_2B/k_BT_2)\big)$ on the field.

\item{(3)} Spin 2 still possesses information $(\tilde S_2
-S_1)/k_B{\rm ln}2$ about spin 1, which can be
converted into work by flipping spin 2 iff spin 1 is in the
state $|\uparrow\rangle_1$, thereby performing
work $p_2(\uparrow) 2\mu_2B {\rm tanh}(\mu_1B/k_BT_2)$
on the field.

\noindent It is straightforward to verify that after these
three conditional spin flips, spin 1 has probabilities
$p'_1(i) = p_2(i)$ while spin 2 has probabilities
$p'_2(i) = p_1(i)$.  That is, the sequence of pulses has
`swapped' the information in spin 1 with the information in
spin 2.$^{12}$  As a result, $S'_1=S_2$, $S'_2=S_1$,
and the new energies of the spins are
$E'_1= -\mu_1B {\rm tanh}(\mu_2 B/ k_B T_2)$ and
$E'_2= -\mu_2B {\rm tanh}(\mu_1 B/ k_B T_1)$.
The total amount of work done by the spins
on the electromagnetic field is
$$\eqalign{W=&-(E'_1+E'_2 - E_1-E_2)\cr
 =& -(\mu_1-\mu_2) B \big({\rm
tanh}(\mu_1 B/k_B T_1) - {\rm tanh}(\mu_2 B/k_B T_2)\big).}
 \eqno(6)$$
\noindent This formula for the work done depends only on
conservation of energy and does not depend on
the specific set of coherent pulses that are used to
`swap' the spins.
Equation (6) shows that $W>0$ iff either
$\mu_1>\mu_2, \allowbreak \mu_1/T_1<\mu_2/T_2$ or
$\mu_1<\mu_2, \allowbreak \mu_1/T_1>\mu_2/T_2$.
If $T_1=T_2$, $W$ is zero or negative: no work
can be extracted from the spins at equilibrium.  The
device cannot function as a {\it perpetuum mobile} of
the second kind.  The cycle
can be completed by letting the spins
re-equilibrate with their respective reservoirs.
Each time steps (1-3) are
repeated, heat $Q_{\rm in}=E_1-E'_1$ flows from reservoir 1
to spin 1
and heat $Q_{\rm out}= E_2'-E_2$ flows from spin 2
into reservoir 2.  The efficiency of this cycle
is $W/Q_{\rm in} = 1- \mu_2/\mu_1 < 1-T_2/T_1 = \varepsilon_C$:
when the spins equilibrate with their respective reservoirs,
heat flows but no work is done.

The following steps can be added to the cycle to allow
the spins to re-equilibrate isentropically:

\bigskip\noindent (5) Return spin 1 to its original state:

\item (i) Take the spin out of `contact' with its reservoir
by varying the frequency of the reservoir modes as above.

\item(ii) Alter the quasi-static field from
$B\rightarrow B_1 = B T_1/T_2$ adiabatically, with no
heat flowing between spin and reservoir.

\item(iii)
Gradually change the field from $B_1\rightarrow B$ keeping
the spin in `contact' with the reservoir at temperature
$T_1$ so that heat
flows isentropically between the spin and the reservoir.

\bigskip\noindent During this process, entropy
$S_2-S_1$ flows from the spin to the reservoir,
while the spin does work $E_1 - E'_1 -T_1(S_2-S_1)$
on the field.

\bigskip\noindent (6) Return spin 2 to its original state
by the analogous set of steps.

The total work done by the spins
on the electromagnetic field throughout the cycle is
$$W_C= (T_1-T_2)(S_1-S_2)
\quad. \eqno(7)$$
\noindent With the added steps (5-6)
performs a Carnot cycle and
in principle operates at the Carnot efficiency $1-T_2/T_1$.
In practice, of course,
the steps that go into operation of the such an engine
will be neither adiabatic nor isentropic, leading
to an actual efficiency below the
Carnot efficiency.

\bigskip\noindent{\bf 3. Thermodynamic cost of quantum
measurement and decoherence}

So far, although the demon has functioned within the laws
of quantum mechanics, quantum measurement and quantum
information have not entered in any fundamental way.
Now that the thermodynamics of the demon have been
elucidated, however, it is possible to quantify precisely
the effects both of measurement and of the introduction of
`new' quantum information on the demon's thermodynamic
efficiency.  The simple quantum information-processing
engine of the previous section can in principle be
operated at the Carnot efficiency: as will now be shown,
when the engine introduces
`new' information by a process of measurement or
decoherence, it cannot be operated even in principle (let alone
in practice) above the lower, quantum efficiency
$\varepsilon_Q$ of equation (4).

To isolate the effects of quantum information,
first consider the simple model of section 1 above, in which
spin 1 is initially in the state
$|\rightarrow\rangle_1 = 1/{\sqrt 2}( |\uparrow\rangle_1
+|\downarrow\rangle_1)$.  This state has non-minimum free energy
which is available for immediate conversion into work: simply apply a
$\pi/2$ pulse to rotate spin 1 into the state
$|\downarrow\rangle_1$, adding energy
$\hbar\omega_2/2=\mu_1 B$ to the
oscillating field in the process.
Suppose, however, that
instead of extracting this energy directly, the demon
operates in information-gathering mode
as above, using magnetic resonance techniques
to correlate the state of spin 2 with the state of spin 1
(cf. reference 5).  Suppose spin 2 is initially in the state
$|\downarrow\rangle_2$.$^{18}$
In this case, coherently flipping spin 2 iff spin 1 is in
the state $|\uparrow\rangle_1$ results in the state,
$1/{\sqrt 2}(|\uparrow\rangle_1|\uparrow\rangle_2 +
|\downarrow\rangle_1 |\downarrow\rangle_2)$, a quantum
`entangled' state in which the state of spin 2 is perfectly
correlated with the state of spin 1.  Continuing
the energy extraction process by flipping spin 1
iff spin 2 is in the state $|\uparrow\rangle_2$
as before allows on average an amount of
energy $(\mu_1-\mu_2) B$ to be extracted from the spin.  The
resulting state of the spins is
$1/{\sqrt 2}|\downarrow\rangle_1|(|\uparrow\rangle_2+
|\downarrow\rangle_2) = 1/{\sqrt 2}|\downarrow\rangle_1
|\rightarrow\rangle_2$.  Up
until this point, no extra thermodynamic cost
has been incurred.  Indeed,
since the conditional spin flipping occurs coherently, the
process can be reversed by repeating the steps in reverse
order to return to the original state $|\rightarrow\rangle_1$,
with a net energy and entropy change of zero.

When is the cost of quantum measurement realized?
When decoherence occurs.  In the original cycle,
decoherence takes place when spin 2 is put
in contact with the reservoir to `erase' it:$^{13-14}$
the exchange of energy between the spin and the reservoir
is an incoherent process during which the pure
state $|\rightarrow\rangle_2=1/{\sqrt 2}(|\uparrow\rangle_2 +
|\downarrow\rangle_2)$ goes to the mixed state described by
the density matrix $(1/2)(|\uparrow\rangle_2\langle
\uparrow| + |\downarrow\rangle_2 \langle \downarrow|)$.
The time scale for this process of decoherence is equal to
the spin dephasing time $T_2^*$ and is typically
much faster than the time scale for the transfer of energy.$^{14}$
In effect, interaction with the reservoir turns the process by
which spin 2 coherently acquires quantum information
about spin 1 into an decoherent process of measurement,
during which a bit of `new' information is created.
This bit corresponds to an increase in entropy
$k_B{\rm ln} 2$.  During the process of erasure, this
entropy is transferred from spin 2 to the
low-temperature reservoir, in accordance with Landauer's
principle.

By decohering and effectively measuring the spin,
the demon has increased
entropy and introduced thermodynamic inefficiency.
The amount of inefficiency can be quantified precisely
by going to the Carnot cycle model of the demon above.
In general,
the state of spin $1$ is described initially by a
density matrix
$$\rho'_1 = p_1(\uparrow') |\uparrow'\rangle_1
\langle\uparrow'| + p_1(\downarrow') |\downarrow'\rangle_1
\langle\downarrow'|,\eqno(8)$$
\noindent where $|\uparrow'\rangle$ and
$|\downarrow'\rangle$ are spin states along an axis at some
angle $\theta$ from the $z$-axis.  Without loss of generality,
$T_1$ and $B$ can be taken to be such that
$p_1(\uparrow') = e^{-\mu_1B/k_BT_1}/Z_1$,
$p_1(\downarrow) = e^{\mu_1B/k_BT_1}/Z_1$.
This state is not at equilibrium, and possesses free
energy that can be extracted by applying a tipping pulse
that rotates the spin by $\theta$ and takes
$|\uparrow'\rangle\longrightarrow |\uparrow\rangle$ and
$|\downarrow'\rangle\longrightarrow |\downarrow\rangle$.
The amount of work extracted is
$$W^*=E_1^*-E_1 \eqno(9)$$
\noindent where
$$E_1=\mu_1B\big(p_1(\uparrow)-p_1(\downarrow)\big),
E^*_1=\mu_1B\big(p^*_1(\uparrow)-p^*_1(\downarrow)\big)
\eqno(10)$$
\noindent and
$$p_1^*(\uparrow) = p_1(\uparrow){\rm cos}^2\theta +
p_1(\downarrow){\rm sin}^2\theta, \quad
p_1^*(\downarrow) = p_1(\downarrow){\rm cos}^2\theta +
p_1(\uparrow){\rm sin}^2\theta \eqno(11)$$
Running the engine
through a Carnot cycle by steps (1-5) above then
extracts work $(T_1-T_2)(S_1-S_2)$.  This process
extracts the free energy of the spin isentropically,
without increasing entropy.

Inefficiency due to measurement arises when instead of
first applying the tipping pulse to extract spin 1's
free energy, one simply operates the engine cyclically
as before.  The steps are as above: three
conditional flips `swap' the states of spin 1 and spin 2,
so that spin 1 is in the state $\rho_2$ and spin 2 is
in the state $\rho'_2=\rho'_1$.  The interaction with the heat
reservoir then decoheres spin 2, destroying
the off-diagonal terms in the density matrix so that
$\rho'_2 \rightarrow p^*_1(\uparrow) |\uparrow\rangle
\langle\uparrow| + p^*_1 |\downarrow\rangle \langle \downarrow|$
with entropy $S^*_1 = -k_B\sum_{i=\uparrow,\downarrow}
p^*_1(i){\rm ln} p^*_1(i)$.  $(S^*_1-S_1)/k_B{\rm ln}2$
is the `extra' information introduced by decoherence.
The entropy $S^*_1-S_2$
that flows out to reservoir 2 is greater than the
entropy $S_1-S_2$ that flowed in from reservoir 1.
The total amount of work done is $k_BT_1(S_1-S_2) -
T_2(S^*_1-S_2)+W^*$, $T_2(S^*_1-S_1)$ less than the
work $(T_1-T_2)(S_1-S_2)+ W^*$
done by simply undoing the tipping pulse and
operating the engine as before.  The overall efficency
with decoherence and measurement included is
$$
\varepsilon_Q = 1-
T_2(S^*_1-S_2)/T_1(S_1-S_2) \leq \varepsilon_C,\eqno(12)$$
\noindent in accordance with equation (4) above.
The extra information introduced by quantum measurement
and decoherence has decreased the efficiency of the
demon.  The quantum efficiency $\varepsilon_Q$ rather
than the Carnot efficiency $\varepsilon_C$ provides the
upper limit to the maximum efficiency of such an engine.

\bigskip\noindent{\bf 4. Conclusion}

This paper presented general arguments and specific
models that show how quantum measurement and decoherence
decrease the efficiency of heat engines.
The systems discussed are experimentally realizable
examples of Zurek's quantum Szilard engine
{\it gedankenexperiment}.$^5$
The techniques given here for exploiting
quantum information to perform work can be
extended in a variety of ways.
If operated in the regime where $\mu_1/T_1>\mu_2/T_2$,
the demon functions as a refrigerator, using the information
gained with the help of work from the electromagnetic
field to pump heat from the reservoir at temperature $T_2$
to the reservoir at temperature $T_1$.$^{19-20}$  Once again,
quantum measurement and decoherence introduce inefficiency
in the operation of such a `Maxwell's fridge.'
The use of magnetic resonance techniques to describe
a quantum Maxwell's demon was for the sake of
convenience of exposition and potential
experimental realizability: many
other quantum systems could be suitable for performing
the heat--information--energy conversion described
above.  At bottom, a quantum `demon' consists of
nothing more than an interaction between two quantum systems
that allows the controlled transfer of information
from one to the other.
In particular, any system that can provide
the coherent quantum logic operation controlled-$NOT$
that flips one quantum bit conditioned on the
state of another could form the basis for an
information-processing quantum heat engine.$^{21-22}$

\vfill
\noindent{\it Acknowledgements:} This paper originated in
a series of discussions with Hermann Haus, who supplied
the crucial insight that Landauer's principle
implies that quantum demons suffered from additional sources
of inefficiency.  Without these discussions, and without
Professor Haus's insistence that the author present a
formal treatment of a quantum demon, this
paper would not have been written.
\eject
\centerline{\it References}
\bigskip

\noindent 1. J.C. Maxwell, {\it Theory of Heat}, Appleton, London,
1871.

\noindent 2. H.S. Leff and A.F. Rex, {\it Maxwell's Demon:
Entropy, Information, Computing,} Princeton University Press,
Princeton, 1990.

\noindent 3. See, e.g., C.H. Bennett, {\it Sci. Am.} {\bf 257},
108 (1987); W.H. Zurek, {\it Nature} {\bf 341}, 119 (1989).

\noindent 4. R. Landauer, {\it IBM J. Res. Dev.} {\bf 5},
183 (1961).

\noindent 5. W.H. Zurek, Leff and Rex {\it op. cit.}, 249-259,
reprinted from {\it Frontiers of Nonequilibrium Statistical
Physics}, G.T. Moore, M.O. Scully, eds., Plenum Press, New
York, 151-161.

\noindent 6. S. Lloyd, {\it Physical Review} {\bf A} 39,
5378 (1989).

\noindent 7. G.J. Milburn, `An atom-optical Maxwell demon,'
University of Queensland preprint, 1995.

\noindent 8. H. Haus, private communication.

\noindent 9. C.P. Slichter, {\it Principles of Magnetic
Resonance}, third edition (Springer-Verlag, New York,
1990).  A $\pi$ pulse is a transversely polarized oscillatory pulse
with integrated intensity $\hbar^{-1} \int \mu H(t) = \pi$, where
$H(t)$ is the envelope function of the oscillatory field.

\noindent 10. O.R. Ernst, G. Bodenhausen, A. Wokaun,
{\it Principles of Nuclear Magnetic Resonance in One and
Two Dimensions,} Oxford University Press, Oxford, 1987.
There are a number of ways to flip one spin coherently
conditioned on the state of another.  For example,
consider a second spin that interacts with the first in an
Ising-like fashion, so that the Hamiltonian for the two
spins is $2\mu_1B\sigma^1_z + 2\mu_2B\sigma^2_z + \hbar
\gamma \sigma^1_z\sigma^2_z$, where $\mu_1=\mu$ and $\mu_2$
are the magnetic dipole moments for the two spins and
$\gamma$ is a coupling constant.  To perform a
controlled-$NOT$,

\item{(i)} Apply a $\pi/2$
pulse with frequency $\omega_2= 2\mu_2B/\hbar$ and
width $<< 2|\mu_1-\mu_2|$ and $>2\gamma$.

\item{(ii)} Wait for time $\pi/2\gamma$.

\item{(iii)} Apply a second $\pi/2$ pulse with a phase
delay of $3\pi/2$ from the first pulse.

\noindent Step (i) rotates spin 2 by $\pi/2$, step (ii) allows spin 2
to acquire a phase of $\pm \pi/2$ conditioned on whether
spin 1 is in the state $|\uparrow\rangle$ or
$|\downarrow\rangle$, and step (iii) either rotates spin 2 back
to its original state if spin 1 $= |\downarrow\rangle$
or rotates spin 2 to an angle of $\pi$ from its original
state is spin 1 $=|\uparrow\rangle$.
Another way to flip spin 2 iff spin 1 $=|\uparrow\rangle$
is to apply a highly selective
$\pi$ pulse with frequency $\omega_2$ and
width $<<2\gamma$.  Spin 1 is off-resonance and does nothing,
while spin 2 is on-resonance and flips iff
spin 1 $=|\uparrow\rangle$.

\noindent 11. S. Lloyd, W.H. Zurek, {\it J. Stat. Phys.}
{\bf 62}, 1991.

\noindent 12. See, e.g., J.A. Wheeler and W.H. Zurek,
{\it Quantum Theory and Measurement}, Princeton University
Press, Princeton, 1983.

\noindent 13. M. Gell-Mann, J. Hartle, in {\it Complexity,
Entropy, and the Physics of Information}, W.H. Zurek ed.,
Santa Fe Institute Studies in the Sciences of Complexity
{\bf VIII}, Addison-Wesley, Redwood City, 1990.

\noindent 14. W.H. Zurek {\it Physics Today} {\bf 44},
36, (1991); {\it Phys. Rev. D} {\bf 24}, 1516 (1981);
{\it Phys. Rev. D} {\bf 26}, 1516 (1981).

\noindent 15. D. Divincenzo, {\it Science} {\bf 270}, 255 (1995).

\noindent 16. D. Deutsch, A. Ekert, R. Jozsa, {\it Phys. Rev.
Lett.} {\bf 74}, 4083 (1995).

\noindent 17. T. Sleator, H. Weinfurter {\it Phys. Rev. Lett.}
{\bf 74}, 4087 (1995).

\noindent 18. In principle,
a slight extension of the cycle allows spin 2 to
be placed in the state $|\downarrow\rangle_2$ at
the start of the Carnot cycle.  Starting
with spin 2 at equilibrium with reservoir 2 at temperature
$T_2$ as above, gradually raise the external field
$B\rightarrow B'>> k_B T_2/\mu_2$ while keeping spin 2
in contact with the reservoir, allowing heat $T_2(k_B
{\rm ln} 2 - S_2)$ to flow
isentropically from the spin to the reservoir: spin 2
is now in the state $|\downarrow\rangle_2$ with high
probability.  Now adiabatically lower the field from
$B'\rightarrow B$.  (Spin 1 should be kept
out of contact with its reservoir throughout this
process.)  Spin 2 now begins the cycle in the
state $|\downarrow\rangle_2$.  In practice, of
course, magnetic fields high enough to perform
this extension are hard to come by.

\noindent 18. For an example of a quantum-optical refrigerator,
see R.I. Epstein, M.I. Buchwald, B.C. Edwards, T.R. Gosnell,
C.E. Mungan, {\it Nature} {\bf 377}, 500 (1995).  Inefficiency
due to decoherence could be observed in their system by applying
intense laser pulses to `tip' the state of the ions during
the operation of the refrigerator.

\noindent 19. J. Smith, `Information Refrigeration,' MIT
Physics and Media Group Technical Report, 1995.

\noindent 20. Q.A. Turchette, C.J. Hood, W. Lange, H. Mabuchi,
H.J. Kimble, {\it Phys. Rev. Lett.} {\bf 75}, 4710 (1995).

\noindent 21. C. Monroe, D. M. Meekhof, B.E. King, W.M. Itano,
D.J. Wineland, {\it Phys. Rev. Lett.} {\bf 75}, 4714 (1995).
The sideband pumping technique used in this reference to
cool the vibrational motion of ions in an ion trap is
another experimentally realized example of a quantum
`Maxwell's fridge.'  Entropy and energy are pumped out of
the translational motion of the ion into hyperfine states
of the ion, whence they are transferred to the environment
as incoherent visible light.

\vfill\eject\end